\documentclass[aps,prd,twocolumn,showpacs,superscriptaddress]{revtex4}
\usepackage{graphicx}
\usepackage{epsfig}

\usepackage{amssymb}
\usepackage{abrev-jour-long}
\usepackage[hidelinks]{hyperref}
\usepackage{layouts}
\usepackage{color}
    \definecolor{Blue}{rgb}{0.0,0.0,1.0}
    \definecolor{Red}{rgb}{1.0,0.0,0.0}
    \definecolor{Green}{rgb}{0.0,1.0,0.0}
    \definecolor{MyRed}{rgb}{0.9,0.0,0.1}

\begin{document}

\title[]{Curvature dependence of relativistic epicyclic frequencies in static, axially symmetric spacetimes}

\author{Ronaldo S. S. Vieira}\email[]{rss.vieira@usp.br}
\affiliation{Instituto de Astronomia, Geof\'{i}sica e Ci\^encias 
Atmosf\'{e}ricas, Universidade de S\~{a}o Paulo, 05508-090, 
S\~{a}o Paulo, SP, Brazil}    
\affiliation{Copernicus Astronomical Center, ul. Bartycka
  18, PL-00-716, Warszawa, Poland} 
\affiliation{Institute of Physics,
  Faculty of Philosophy and Science, Silesian University in Opava,
  Bezru{\v c}ovo n{\'a}m. 13, CZ-74601 Opava, Czech Republic} 

\author{W\l odek Klu\'zniak}\email[]{wlodek@camk.edu.pl}
\affiliation{Copernicus Astronomical Center, ul. Bartycka 18,
               PL-00-716, Warszawa, Poland}  
\affiliation{Institute of Physics, Faculty of Philosophy and Science, Silesian
University in Opava, Bezru{\v c}ovo n{\'a}m. 13, CZ-74601 Opava, 
Czech Republic} 

\author{Marek Abramowicz}\email[]{marek.abramowicz@physics.gu.se}
\affiliation{Copernicus Astronomical Center, ul. Bartycka 18,
               PL-00-716, Warszawa, Poland} 
\affiliation{Institute of Physics, Faculty of Philosophy and Science, Silesian
University in Opava, Bezru{\v c}ovo n{\'a}m. 13, CZ-74601 Opava, 
Czech Republic} 
\affiliation{Physics Department, Gothenburg University,
               SE-412-96 G{\"o}teborg, Sweden} 

\date{\today}

%
\begin{abstract}
The sum of squared epicyclic frequencies of nearly circular motion ($\omega_r^2+\omega_\theta^2$) in axially symmetric configurations of Newtonian gravity is known to depend both on the matter density and on the angular velocity profile of circular orbits. It was recently found that this sum goes to zero at the photon 
orbits of Schwarzschild and Kerr spacetimes.
However, these are the only relativistic configurations for which such result exists in the literature.
Here, we extend the above formalism in order to describe the analogous relation for geodesic motion in arbitrary
static, axially symmetric, asymptotically flat solutions of general relativity. The sum of squared epicyclic frequencies is found to vanish 
at photon radii of vacuum solutions. In the presence of matter, we obtain that $\omega_r^2+\omega_\theta^2>0$ for perturbed timelike circular 
geodesics on the equatorial plane if the strong energy condition holds for the matter-energy fluid of spacetime; in vacuum, the allowed region for timelike circular geodesic motion is characterized by the inequality above.
The results presented here may be of use to shed light on general issues concerning the stability of circular orbits once they approach 
photon radii, mainly the ones corresponding to stable photon motion. 
\end{abstract}


\maketitle



\section{Introduction}

General relativity (GR) introduces many additional features which are not present in Newtonian gravity. For instance, the radial and vertical epicyclic frequencies are not equal in spherically symmetric spacetimes, a fact first pointed out by \cite{shirokov1973GRG}. The difference between these frequencies in Schwarzschild spacetime generates an innermost stable circular orbit (at which the radial epicyclic frequency vanishes), which determines the inner rim of thin accretion discs around black holes. Recently, 
Amsterdamski et al. \cite{amsterdamskiEtal2002AA}
found outside Newtonian Maclaurin spheroids a minimum radius inside which there are no circular orbits, a phenomenon which was believed to exist only in GR. Related to that, Klu\'zniak and Rosi\'nska \cite{kluzniakRosinska2013MNRAS} presented a relation between the sum of the squared 
epicyclic frequencies $\omega_r$ (radial) and $\omega_\theta$ (vertical), the density $\rho$ of the background matter and the orbital frequency $\Omega$ of the original circular orbit (see also \cite{binneytremaineGD}),
  \begin{equation}\label{eq:w2w2Newt}
   \omega_r^2+\omega_\theta^2=2\Omega^2+4\pi G\rho.
  \end{equation}
They also found that the (formal) expression for the sum of the squared 	epicyclic frequencies goes to zero at photon orbits of Kerr spacetime \cite{kluzniakRosinska2013MNRAS}.

The purpose of the present paper is to extend Eq.~(\ref{eq:w2w2Newt}),
valid for Newtonian gravity, to equatorial circular geodesics in static, axially symmetric spacetimes.
In Section \ref{sec:spacetime} we summarize the formalism to obtain the epicyclic frequencies for nearly circular motion in static, axially symmetric spacetimes, following the derivation by \cite{abramowiczKluzniak2005ApSS}. Section \ref{sec:curvature} contains the results of our work, 
namely the relation between the sum \mbox{$\omega_r^2+\omega_\theta^2$} and the Ricci tensor, as well as the particular case of vacuum GR spacetimes
and the relation between the allowed regions for circular geodesics (in terms of the sign of \mbox{$\omega_r^2+\omega_\theta^2$}) and the strong energy condition for the spacetime energy content. We present our conclusions in Section \ref{sec:conclusions}.

\section{Static, axially symmetric spacetimes}\label{sec:spacetime}

Let \mbox{$\eta=\partial/\partial t$} and \mbox{$\xi=\partial/\partial\varphi$} be the timelike and azimuthal Killing vector fields of a static, axially symmetric spacetime. The line element can be written, in a coordinate system adapted to these Killing vector fields, as
  \begin{equation}\label{eq:metric}
   ds^2 = g_{tt}dt^2 + g_{\varphi\varphi}d\varphi^2 + g_{rr}dr^2 + g_{\theta\theta}d\theta^2.
  \end{equation}
The metric coefficients depend only on $R$ and $\theta$. The metric signature is taken as \mbox{(+ -- -- --)}
and the spacetime is assumed asymptotically flat, so \mbox{$g_{\mu\nu}\to\eta_{\mu\nu}$} (Minkowski metric) at spatial infinity.
We also impose equatorial-plane symmetry. In particular, \mbox{$g_{\mu\nu,\theta}=0$} on the equatorial plane (defined by \mbox{$\theta=\pi/2$}). We must stress that the formalism presented in this paper is valid for smooth metrics; the vertical stability criteria for circular orbits in razor-thin disks were analyzed in \cite{vieiraRamosCaro2016CeMDA, vieiraRamoscaroSaa2016PRD}.

We adopt the conventions of \cite{abramowiczKluzniak2005ApSS}. In particular, 
the specific angular momentum \mbox{$\ell=-u_\varphi/u_t$} of timelike circular geodesics on the equatorial 
plane is given by 
\mbox{$\ell^2(r)= - g^{tt}_{ \ ,r}/g^{\varphi\varphi}_{\ \ ,r}$}, 
where $u^\mu$ is the 4-velocity of the geodesic.   
The corresponding angular velocity $\Omega$ is given by
\mbox{$\Omega^2(r)=-g_{tt,r}/g_{\varphi\varphi,r}$}
and these quantities are related by
  \begin{equation}
   \ell^2(r)=\tilde{r}^4\Omega^2(r),
  \end{equation}
where
  \begin{equation}
   \tilde{r}^2=-\frac{g_{\varphi\varphi}}{g_{tt}}
  \end{equation}
is the radial coordinate in optical geometry \cite{abramowicz1990MNRAS, katkaVieiraEtal2015GRG}.
The radial ($\omega_r$) and vertical ($\omega_\theta$) epicyclic frequencies of nearly circular orbits, measured at infinity, are given by \cite{abramowiczKluzniak2005ApSS}
  \begin{equation}\label{eq:omega-r}
   \omega_r^2=-\frac{(g_{tt})^2}{2|g_{rr}|}\Big[g^{tt}_{\ \ ,rr} + \ell^2(r)g^{\varphi\varphi}_{\ \ ,rr}\Big],
  \end{equation}
  \begin{equation}\label{eq:omega-theta}
   \omega_\theta^2=-\frac{(g_{tt})^2}{2|g_{\theta\theta}|}
    \Big[g^{tt}_{\ \ ,\theta\theta} + \ell^2(r)g^{\varphi\varphi}_{\ \ ,\theta\theta}\Big].
  \end{equation}
In terms of $d\ell^2/dr$, the frequency $\omega_r$ can be written as
  \begin{equation}\label{omegar2usado}
   \omega_r^2= \frac{1}{2\tilde{r}^4 g_{rr}}g_{\varphi\varphi,r}\frac{d\ell^2(r)}{dr}.
  \end{equation}
  
We may write $d\ell^2/dr$ in terms of $d\tilde{r}^2/dr$ and $d\Omega^2(r)/dr$: 
  \begin{equation}\label{dell2dr}
   \frac{d\ell^2(r)}{dr}=2\tilde{r}^2\Omega^2(r)\frac{d\tilde{r}^2}{dr} + \tilde{r}^4\frac{d\Omega^2(r)}{dr}.
  \end{equation}
Combining Eqs.~(\ref{omegar2usado}) and (\ref{dell2dr}),
we obtain
  \begin{equation}\label{eq:omega-r2-rtilde}
   \omega_r^2=\frac{g_{\varphi\varphi,r}}{g_{rr}}\Omega^2(r)\frac{1}{\tilde{r}^2}\frac{d\tilde{r}^2}{dr} + 
    \frac{g_{\varphi\varphi,r}}{2 g_{rr}}\frac{d\Omega^2(r)}{dr}.
  \end{equation}

\section{Relation between epicyclic frequencies and curvature}\label{sec:curvature}
We adopt the following conventions for the Riemann ($R_{\alpha\beta\mu}^{\ \ \ \ \nu}$) and Ricci ($R_{\mu\nu}$) tensors: 
  \begin{equation}
   R_{\alpha\beta\mu}^{\ \ \ \ \nu} = \partial_\alpha \Gamma^\nu_{\beta\mu} - \partial_\beta \Gamma^\nu_{\alpha\mu} + \Gamma^\sigma_{\beta\mu}\Gamma^\nu_{\alpha\sigma} - \Gamma^\sigma_{\alpha\mu}\Gamma^\nu_{\beta\sigma}
  \end{equation}  
  \begin{equation}
  R_{\mu\nu}=R_{\alpha\mu\nu}^{\ \ \ \ \alpha},
  \end{equation}
where the Christoffel symbols $\Gamma^\mu_{\nu\alpha}$ are given by
    \begin{equation}
   \Gamma^\mu_{\nu\alpha}=\frac{1}{2}g^{\mu\sigma}
   \big(g_{\sigma\alpha,\,\nu}+g_{\sigma\nu,\,\alpha}-g_{\nu\alpha,\,\sigma}\big).
  \end{equation}   
According to the Newtonian relation~(\ref{eq:w2w2Newt}), we expect that 
the relativistic expression for \mbox{$\omega_r^2+\omega_\theta^2$} will depend on some invariant function of $R_{\mu\nu}$, which must reduce to $4\pi G\rho$ in the Newtonian limit. 
This scalar function of $R_{\mu\nu}$ would give the dependence of \mbox{$\omega_r^2+\omega_\theta^2$} on the matter--energy content of spacetime.
Also, we know from the Schwarzschild case \cite{kluzniakRosinska2013MNRAS} that the squared sum of epicyclic frequencies
formally vanishes at the radius of the circular photon orbit, hereafter called \textit{photon radius} $r_{\rm ph}$. 
We thus expect that the sum \mbox{$\omega_r^2+\omega_\theta^2$} would involve two terms: the first term being related to the local matter-energy content of the spacetime (i.e. the Ricci tensor)
and the second term vanishing at the photon radius. This second term must reduce to $2\Omega^2$ in the Newtonian limit, according to Eq.~(\ref{eq:w2w2Newt}).
  
Equatorial-plane symmetry implies that the Ricci tensor $R_{\mu\nu}$ is diagonal at $\theta=\pi/2$.
Since the epicyclic frequencies do not depend on partial derivatives of $g_{rr}$ and $g_{\theta\theta}$, these terms should not appear 
in the corresponding combination of Ricci tensor coefficients. On the other hand,  \mbox{$R_{tt}=4\pi G\rho$} in the Newtonian limit,
which implies that this term should appear in the expression for \mbox{$\omega_r^2+\omega_\theta^2$}.
If we look for linear combinations of the coefficients $R_{\mu\nu}$, the curvature-like contribution 
must be of the form \mbox{$R_{tt}+\Omega^2(r)R_{\varphi\varphi}$} in order to 
cancel $g_{rr}$ and $g_{\theta\theta}$ derivatives.
We see next that this is indeed the correct expression; in fact we have on the equatorial plane, by direct computation,
  \begin{eqnarray}\label{eq:RttRpp}
   R_{tt}+ \Omega^2(r)R_{\varphi\varphi} &=& \Big(\partial_r\Gamma^r_{tt} + \Omega^2 \partial_r\Gamma^r_{\varphi\varphi}\Big) +\nonumber\\
   & & + \Big(\partial_\theta\Gamma^\theta_{tt} + \Omega^2 \partial_\theta\Gamma^\theta_{\varphi\varphi}\Big)  \\
  & & + 2 \Gamma^r_{tt}\Big(\Gamma^\varphi_{r \varphi} - \Gamma^t_{r t}\Big). \nonumber
  \end{eqnarray}
Equation~(\ref{eq:omega-theta})
for $\omega_\theta^2$ allows us to write
  \begin{equation}\label{eq:Gamma-theta}
   \omega_\theta^2 = \partial_\theta\Gamma^\theta_{tt} + \Omega^2 \partial_\theta\Gamma^\theta_{\varphi\varphi}.
  \end{equation}
The other two terms of (\ref{eq:RttRpp}) can be written as 
  \begin{equation}\label{eq:Gamma-r}
   \partial_r\Gamma^r_{tt} + \Omega^2 \partial_r\Gamma^r_{\varphi\varphi} = 
   \frac{g_{\varphi\varphi,r}}{2 g_{rr}}\frac{d\Omega^2(r)}{dr}
  \end{equation}
and 
  \begin{equation}\label{eq:2Gamma-r}
   2 \, \Gamma^r_{tt}\Big(\Gamma^\varphi_{r \varphi} - \Gamma^t_{rt}\Big) = 
    \frac{1}{2}\frac{g_{\varphi\varphi,r}}{g_{rr}}\Omega^2(r)\frac{1}{\tilde{r}^2}\frac{d\tilde{r}^2}{dr}.
  \end{equation}
We can then separate the contribution of $d\tilde{r}^2/dr$ appearing in (\ref{eq:omega-r2-rtilde}) with the help of
Eqs.~(\ref{eq:Gamma-r}) and (\ref{eq:2Gamma-r}) in two terms,
  \begin{eqnarray}\label{eq:omega-r2-Gamma}
   \omega_r^2 &=&  \frac{1}{2}\frac{g_{\varphi\varphi,r}}{g_{rr}}\Omega^2(r)\frac{1}{\tilde{r}^2}\frac{d\tilde{r}^2}{dr} + \nonumber \\
    & & +\bigg[2 \, \Gamma^r_{tt}\Big(\Gamma^\varphi_{r \varphi} - \Gamma^t_{rt}\Big) +  
    \frac{g_{\varphi\varphi,r}}{2 g_{rr}}\frac{d\Omega^2(r)}{dr}\bigg]. 
  \end{eqnarray}

The curvature term \mbox{$R_{tt}+ \Omega^2(r)R_{\varphi\varphi}$} is then easily recognized from Eq.~(\ref{eq:RttRpp})  
in the combination \mbox{$\omega_r^2 + \omega_\theta^2$}, according to Eqs.~(\ref{eq:Gamma-theta}), (\ref{eq:Gamma-r}), (\ref{eq:2Gamma-r}), and (\ref{eq:omega-r2-Gamma}). The final expression is written 
in terms of $R_{\mu\nu}$, $\Omega(r)$ and $\tilde r$ as
  \begin{equation}\label{eq:w2+w2}
   \omega_r^2 + \omega_\theta^2 = \Big[R_{tt}+ \Omega^2(r)R_{\varphi\varphi}\Big] +
    \frac{g_{\varphi\varphi,r}}{2g_{rr}}\Omega^2(r)\frac{1}{\tilde{r}^2}\frac{d\tilde{r}^2}{dr}.
  \end{equation}  
The first term on the right-hand side is zero when the spacetime is Ricci flat, while the second term
vanishes at photon radii (\mbox{$d\tilde{r}^2/dr=0$}, see \cite{vieiraMarekEtal2014PRD}) and when \mbox{$\Omega^2(r)=0$} 
(the same radius at which \mbox{$\ell^2=0$} and the radial acceleration of a static observer is null, because \mbox{$g_{tt,r}=0$}
at this radius). The existence of such radius where \mbox{$\Omega^2(r)=0$} is a characteristic of few spacetimes, for instance the Reissner--Nordstr\"om metric in the naked-singularity regime \cite{puglieseQuevedoRuffini2011PRD} and its generalizations 
\cite{stuchlikHledik2002AcPSl}.
This behavior is also present in the Kehagias-Sfetsos naked singularity spacetime \cite{vieiraMarekEtal2014PRD, katkaVieiraEtal2015GRG} and in 
the no-horizon parameter region of what would be otherwise a regular black hole spacetime \cite{garciaHackmannEta2015JMP, boshkakayev2016PRD}.

Equation~(\ref{eq:w2+w2})
reduces to the Newtonian result [Eq.~(4) of \cite{kluzniakRosinska2013MNRAS}; see also Eq.~(\ref{eq:w2w2Newt})] in the appropriate limit: 
\mbox{$R_{tt}\approx 4\pi G\rho$}, \mbox{$R_{\varphi\varphi}\approx 0$}, \mbox{$\tilde r\approx r$}, \mbox{$g_{rr}\approx -1$},
\mbox{$g_{\varphi\varphi,r}\approx -2r$}.

\subsection{Vacuum (Ricci-flat) spacetimes}
  
\begin{figure}
\begin{center}
  \epsfig{figure=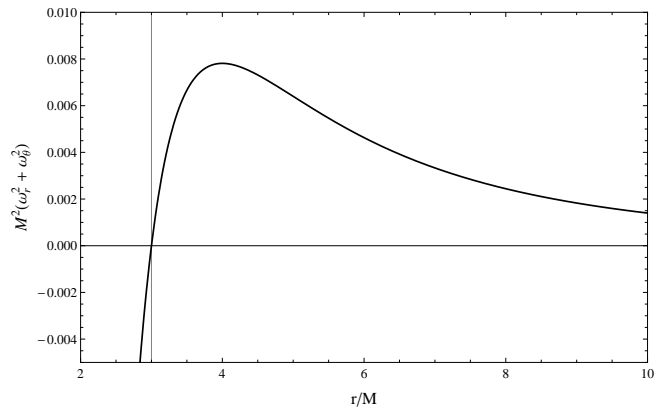,width=0.99\columnwidth ,angle=0}
\caption{The quantity \mbox{$\omega_r^2+\omega_\theta^2$} as a function of the radial coordinate $r$ for the Schwarzschild metric (solid black line). The parameter $M$ is the Schwarzschild mass. As we know from \cite{kluzniakRosinska2013MNRAS}, this quantity goes to zero at the photon radius \mbox{$r_{\rm ph}=3\,M$} (marked by a vertical gray line in the Figure). Our approach extends this result: the Schwarzschild metric is a particular case of a vacuum spacetime in GR. We have \mbox{$\omega_r^2+\omega_\theta^2>0$} for \mbox{$r>r_{\rm ph}$} and \mbox{$\omega_r^2+\omega_\theta^2<0$} for \mbox{$r<r_{\rm ph}$} [see Eq.~(\ref{eq:w2+w2--Ricci-flat})].
}
\label{fig:w2Sch}
\end{center}
\end{figure}

In Ricci-flat spacetimes (vacuum spacetimes for general relativity), Eq.~(\ref{eq:w2+w2}) reduces to
  \begin{equation}\label{eq:w2+w2--Ricci-flat}
   \omega_r^2 + \omega_\theta^2 = 
   \frac{g_{\varphi\varphi,r}}{2g_{rr}}\Omega^2(r)\frac{1}{\tilde{r}^2}\frac{d\tilde{r}^2}{dr};
  \end{equation}  
this expression vanishes at photon radii. This simple statement has tremendous consequences. 
First of all, we must stress that in this case \mbox{$\omega_r^2 + \omega_\theta^2=0$} {\it precisely} at photon radii,
being positive in allowed regions for circular orbits (\mbox{$d \tilde r /dr>0$} and \mbox{$\Omega^2(r)>0$}) and -- formally -- negative in forbiden regions
(\mbox{$d \tilde r /dr<0$} or \mbox{$\Omega^2(r)<0$}). In particular, there are no timelike circular geodesics in the region where \mbox{$\omega_r^2 + \omega_\theta^2\leq 0$}.
This is the proper extension of the Newtonian result \cite{kluzniakRosinska2013MNRAS} with \mbox{$\rho=0$}, Eq.~(\ref{eq:w2w2Newt}), to geodesic motion in 
static, Ricci-flat spacetimes (in particular to test-particle motion in vacuum GR). We exemplify the behavior of \mbox{$\omega_r^2 + \omega_\theta^2$} in vacuum spacetimes by plotting it as a function of radius for the Schwarzschild metric in Fig.~\ref{fig:w2Sch}.

Apart from the degenerate case in which the expressions for both $\omega_r^2$ and $\omega_\theta^2$ are zero at the photon 
radius, the general situation is that one of the squared ``frequencies'' is positive and the other is negative. By continuity, 
there will be a finite region around the photon radius in which one of the squared frequencies is negative. Therefore, 
in Ricci-flat spacetimes there will always be a region of unstable timelike circular geodesics between 
the photon radius and the radius of the closest marginally stable orbit; the region of stability will never reach the photon radius.

This phenomenon also follows from the behavior of $\ell^2(r)$ in the case of unstable photon orbits (in this case timelike circular motion is allowed only for $r>r_{\rm ph}$),
since $\ell^2(r)$ grows as $r$ approaches the photon radius from the right, which corresponds to radial instability of the corresponding circular orbits
\cite{abramowiczKluzniak2005ApSS}. 
Let us remark that the relation between the conserved angular momentum \mbox{$h=-u_\varphi$} and conserved energy \mbox{$E=u_t$} of 
timelike circular geodesics
[Eq.~(11) of \cite{letelier2003PRD}] in static, axially symmetric spacetimes with equatorial-plane symmetry 
allows us to show that $d\ell^2/dr$ and $dh^2/dr$ have the same sign in allowed regions for 
this kind of motion, since \cite{vieiraMarekEtal2014PRD}
  \begin{equation}
   \frac{d\ell^2}{dr}=\frac{1}{E^2}\big[1-v^2\big]\frac{dh^2}{dr},
  \end{equation}
where $v$ is the speed of the particle as measured by a local static observer.
We can also show from these arguments that \cite{vieiraMarekEtal2014PRD, letelier2003PRD}
  \begin{equation}\label{h2}
   h^2(r)=\tilde{r}^4g_{tt,r}\Big(\frac{d\tilde{r}^2}{dr}\Big)^{-1}.
  \end{equation}
  Therefore, since \mbox{$h^2\to\infty$} when $r$ approaches the photon radius, $\ell^2$ grows as 
\mbox{$r\to r_{\rm ph}$}.
Moreover, circular geodesic motion is forbidden if \mbox{$d\tilde{r}^2/dr<0$} 
(the region in $r$ between a stable and an unstable photon orbit), since in this case \mbox{$h^2<0$}.

However, the case of radially stable photon orbits is subtler and more interesting (in this case timelike circular motion is allowed only for $r<r_{\rm ph}$). Although in this situation
$\ell^2(r)$ grows with $r$ until reaching from the left the photon radius, corresponding to \mbox{$\omega_r^2>0$}, the 
fact that the expression for \mbox{$\omega_r^2+\omega_\theta^2$} vanishes at photon orbits means that the value of $\omega_\theta^2$ 
must become negative as the circular orbit approaches the photon orbit.
The orbits become vertically unstable as they approach this photon radius (although they are radially stable because of Rayleigh's criterion \cite{letelier2003PRD}).
This result means that off-equatorial motion has a fundamental importance to the 
stability analysis of circular equatorial geodesic motion in vacuum spacetimes, shedding light on the recent questions 
raised in the literature concerning the behavior of circular geodesics in vacuum multipole solutions of Einstein's equations
\cite{hernandezpastoraHerreraOspino2013PRD}. In particular, for the mentioned multipole solutions, the radial stability analysis
is not sufficient to classify circular geodesic motion in regions near the central compact object when radially stable photon orbits are present.
Therefore, even if circular orbits are radially stable in this region, they will eventually become vertically unstable as 
$r$ grows in the direction of the photon orbit.

The Ricci-flat condition is in fact too stringent. 
It is not necessary that the spacetime is globally Ricci flat; all the above arguments are also valid locally under the 
only requirement that \mbox{$R_{\mu\nu}=0$} in a radial interval containing the photon radius.

\subsection{Background matter and energy conditions}
  
If \mbox{$R_{tt} + \Omega^2(r)R_{\varphi\varphi}\neq0$} at the photon radius, there is not a clear relation anymore between 
the sign of \mbox{$\omega_r^2 + \omega_\theta^2$} and the allowed regions for timelike equatorial circular geodesics. 
In an arbitrary static, axially symmetric spacetime, there is no ``first-principles'' argument which prohibits the 
existence of (unstable) circular timelike geodesics with \mbox{$\omega_r^2 + \omega_\theta^2<0$}. 

\begin{figure}
\begin{center}
  \epsfig{figure=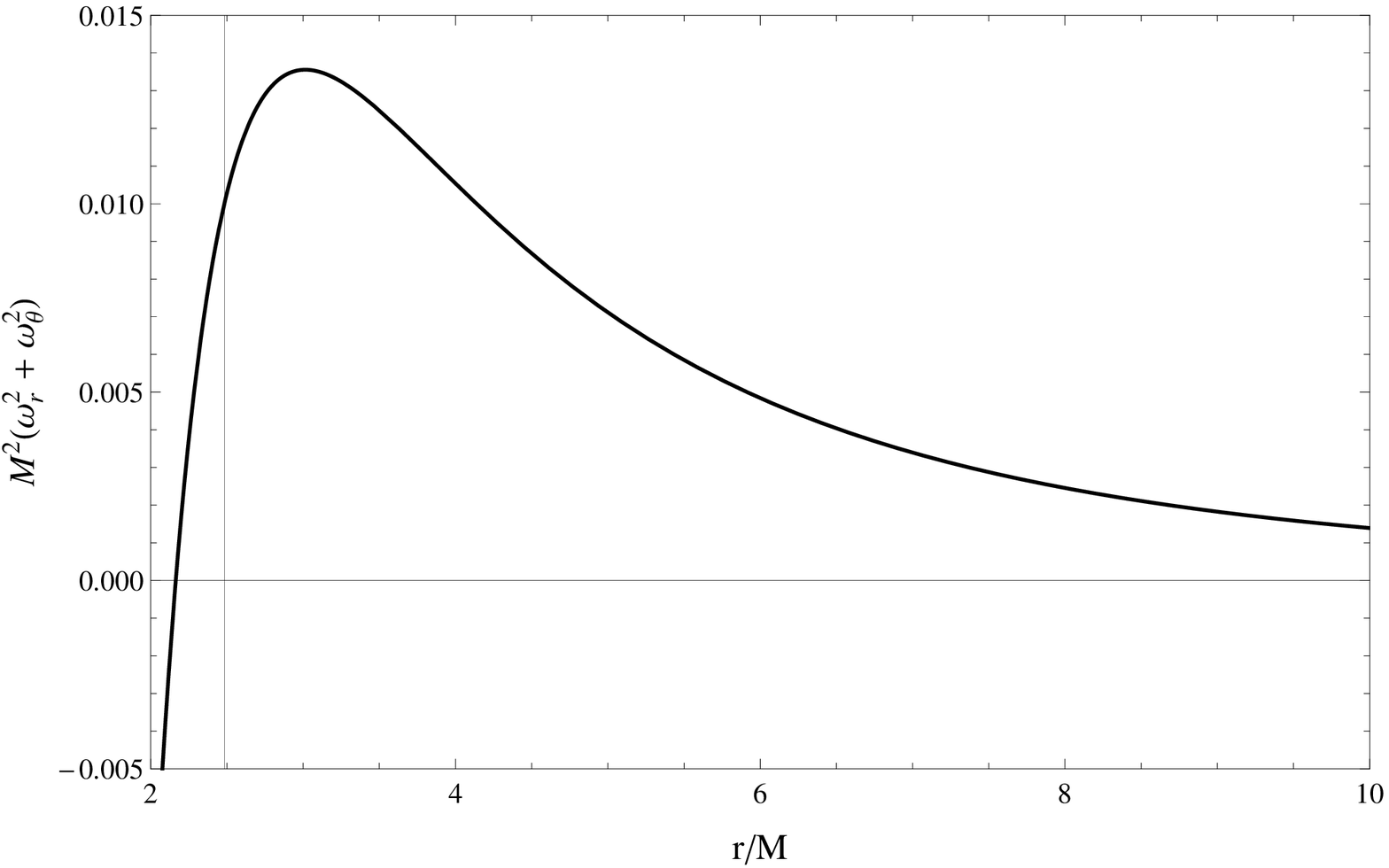,width=0.99\columnwidth ,angle=0}
\caption{The quantity \mbox{$\omega_r^2+\omega_\theta^2$} as a function of the radial coordinate $r$ for the Reissner-Nordstr\"om metric (solid black line). We adopted \mbox{$q=Q/M=0.8$}. The vertical gray line represents the photon radius \mbox{$r_{\rm ph}=2.485\,M$}. We see that \mbox{$\omega_r^2+\omega_\theta^2>0$} for every timelike circular geodesic, as predicted from Eq.~(\ref{eq:w2+w2}) since the Reissner-Nordstr\"om metric satisfies the strong energy condition, Eq.~(\ref{eq:SEC}).
There is also a region inside the photon radius (where circular motion is not allowed) for which the formal expression $\omega_r^2+\omega_\theta^2$ is positive.
}
\label{fig:w2RN}
\end{center}
\end{figure}

We might conclude that this is a purely general relativistic effect, since it does not appear in Newtonian gravity
\cite{kluzniakRosinska2013MNRAS}. This apparent difference comes from the fact that the density of matter is always positive, \mbox{$\rho\geq0$}.
An appropriate generalization of this requirement must be imposed in our case, in order to maintain the well-behaved properties of the energy-momentum tensor.
The established conditions for regular behavior of the matter and energy content of spacetime are the \textit{energy conditions} \cite{waldGR}. 
We find here that the \textit{strong energy condition}
\cite{waldGR} for the background matter,
  \begin{equation}\label{eq:SEC}
   R_{\mu\nu}X^\mu X^\nu \geq 0 \textsl{\ for all timelike vectors \ } X^\mu,
  \end{equation}
implies that all circular timelike geodesics have \mbox{$\omega_r^2 + \omega_\theta^2>0$} for their perturbed motion.  
In fact, writing the tangent unit vector to the aforementioned geodesics as 
  \begin{equation}\label{circularvectorfield}
   u^\alpha=A\big[\eta^\alpha+\Omega(r)\xi^\alpha\big] 
  \end{equation}
with $A=A(r)>0$, we have the equality 
  \begin{equation}\label{eq:Rmunustrong}
   R_{tt} + \Omega^2(r)R_{\varphi\varphi} = \frac{1}{A^2}R_{\mu\nu}u^\mu u^\nu.
  \end{equation}
The ``circular vector field'' $u^\alpha$ is timelike in regions where $\tilde r$ increases with $r$ ($d\tilde r/dr>0$), spacelike
where $\tilde r$ decreases with $r$ ($d\tilde r/dr<0$) and null at photon orbits. 
We have then the following picture: if \mbox{$R_{\mu\nu}u^\mu u^\nu\geq0$} for all timelike circular vector fields of the form
(\ref{circularvectorfield}), then there are no timelike circular geodesics whose expressions for the ``epicyclic frequencies'' satisfy 
\mbox{$\omega_r^2 + \omega_\theta^2\leq0$}. 
Since the strong energy condition (\ref{eq:SEC}) implies the above result, we obtain that \textit{the strong energy condition for the background matter guarantees that $\omega_r^2 + \omega_\theta^2 > 0$ in the allowed region for timelike circular motion}.
As a consequence the strong energy condition is, in our context, the appropriate relativistic extension to the positivity of mass in Newtonian gravity.
The behavior of \mbox{$\omega_r^2 + \omega_\theta^2$} as a function of the radial coordinate is illustrated in Fig.~\ref{fig:w2RN} for the case of the Reissner-Nordstr\"om metric with charge-to-mass parameter \mbox{$q=0.8$}.

Moreover, if \mbox{$R_{tt} + \Omega^2(r)R_{\varphi\varphi}>0$} at the photon radius, we do not have anymore the characterization of the
allowed regions for circular motion in terms of \mbox{$\omega_r^2 + \omega_\theta^2$}, presented in the former Section for Ricci-flat 
spacetimes. In fact, in this case \mbox{$\omega_r^2 + \omega_\theta^2>0$} at the photon radius, and therefore it is possible to 
have an inner region of stability which reaches a radially stable photon orbit (this situation indeed happens for many naked singularity 
spacetimes \mbox{\cite{puglieseQuevedoRuffini2011PRD,  stuchlikHledik2002AcPSl,vieiraMarekEtal2014PRD}} and spacetimes without horizons 
\cite{garciaHackmannEta2015JMP, boshkakayev2016PRD}). 
On the other hand, although $\omega_r^2 + \omega_\theta^2>0$ in the region outside unstable photon orbits, the corresponding circular geodesics are radially unstable according to Rayleigh's criterion.
Therefore, Eq.~(\ref{eq:w2+w2}) does not tell us about the stability of circular orbits without additional assumptions about the spacetime structure.

Precise conditions for the existence of an instability region between the photon radius and the radius of the 
marginally stable orbit can be obtained when spherical symmetry is assumed.
In this case \mbox{$\omega_\theta^2=\Omega^2$}, which is positive in allowed regions for circular 
timelike geodesics. We can therefore write Eq.~(\ref{eq:w2+w2}) as an equation for $\omega_r^2$:
  \begin{equation}\label{wr2spherical}
   \omega_r^2 = \Big[R_{tt}+ \Omega^2(r)\big(R_{\varphi\varphi}-1\big)\Big] +
    \frac{g_{\varphi\varphi,r}}{2g_{rr}}\Omega^2(r)\frac{1}{\tilde{r}^2}\frac{d\tilde{r}^2}{dr}.
  \end{equation}
The condition to guarantee that the stability region reaches the photon orbit is therefore
  \begin{equation}
   R_{tt} + \Omega^2(r)\big(R_{\varphi\varphi} - 1) > 0
  \end{equation}
at the photon radius $r_{\rm ph}$, since from Eq.~(\ref{wr2spherical}) it is equivalent to have \mbox{$\omega_r^2>0$} at the photon radius.
If \mbox{$R_{tt} + \Omega^2(r)\big(R_{\varphi\varphi} - 1) < 0$} at $r_{\rm ph}$, then there will be an instability region between 
$r_{\rm ph}$ and the next marginally stable circular geodesic, whereas if 
\mbox{$R_{tt} + \Omega^2(r)\big(R_{\varphi\varphi} - 1) = 0$} at $r_{\rm ph}$ the criterion is inconclusive.
The above inequality can also be written in terms of spacetime invariants as 
  \begin{equation}
   \Big[R_{\mu\nu}-\xi_\mu \xi_\nu\Big]u^\mu u^\nu>0, 
  \end{equation}
evaluated at the photon radius.

According to Eq.~(\ref{wr2spherical}) and Rayleigh's criterion, we expect  
\mbox{$R_{tt} + \Omega^2(r)\big(R_{\varphi\varphi} - 1) \leq 0$} at radii corresponding to radially unstable photon orbits in spherically symmetric spacetimes. By the same argument, 
the radial region just inside a stable photon orbit corresponds to stable timelike circular geodesics, and thus 
\mbox{$R_{tt} + \Omega^2(r)\big(R_{\varphi\varphi} - 1) \geq 0$} at radii corresponding to radially stable photon orbits.

\section{Conclusions}\label{sec:conclusions}

Circular orbits have been studied since the beginnings of the astrophysical applications of Newtonian gravity and
general relativity. Nevertheless, it was only recently that their qualitative properties in general relativity received proper attention. 
We presented in this paper closed-form expressions for the sum of the squared epicyclic frequencies of perturbed timelike 
equatorial circular geodesics in static, axially symmetric, asymptotically flat spacetimes; these expressions are written in terms of the Ricci tensor and of a quantity which vanishes at photon orbits.
For Ricci-flat spacetimes, the present
framework establishes the existence of an instability region around each photon radius. 
Although this result is a consequence of Rayleigh's criterion near the radius of a radially unstable photon orbit, the same criterion implies that circular geodesics
are radially stable in the inner region near the radius of a radially stable photon orbit. The mentioned region of instability means, in this case, that 
the formula for $\omega_\theta^2$ is negative in this inner region, in the vicinity of the photon radius.
Therefore, although $\omega_r^2$ is positive in this region, 
motion is unstable under off-equatorial perturbations if we get close enough to the radius of the stable photon orbit.
This simple statement has a profound impact on the analysis of circular geodesics in quasi-spherical, multipolar vacuum 
solutions of Einstein's equations \cite{hernandezpastoraHerreraOspino2013PRD}: 
radial perturbations are not sufficient to analyze circular motion 
near an existing inner photon orbit. If off-equatorial perturbations are also considered, these geodesics will eventually become 
unstable before reaching the inner photon radius (and therefore a hypothetic thin accretion disc in this inner region will have a `gap' between its marginally stable circular orbit and the inner photon radius). 

If \mbox{$R_{\mu\nu}\neq 0$}, it is also possible to find a relation between the sign of \mbox{$\omega_r^2+\omega_\theta^2$} and the 
allowed regions for equatorial circular orbits, if the strong energy condition is satisfied for the spacetime matter-energy content.
Namely, in this case \mbox{$\omega_r^2+\omega_\theta^2>0$} in allowed regions for timelike circular geodesics.
Moreover, the formalism presented here is the relativistic generalization of the analogous result in Newtonian
gravity \cite{kluzniakRosinska2013MNRAS, binneytremaineGD} for static, axially symmetric configurations. The dependence of 
\mbox{$\omega_r^2+\omega_\theta^2$} on the Ricci tensor shows the proper generalization of the matter density term in the 
Newtonian equation for our case [see Eqs.~(\ref{eq:w2w2Newt}) and (\ref{eq:w2+w2})], the strong energy condition being the correct relativistic generalization of the positivity of mass (as it was recently found in \cite{vieiraRamoscaroSaa2016PRD} for relativistic razor-thin disks). 
Thus, as in many other qualitative results for timelike geodesic motion in Lorentzian manifolds \cite{waldGR}, 
the results obtained here have a deep relation with energy conditions for the background spacetime.

Modified theories of gravity do not have, necessarily, (\ref{eq:SEC}) as an energy condition; the properties of geodesic motion may not be directly connected to the properties of matter in these theories. 
However, even in these theories the above arguments are valid, since they only depend on the properties of the metric and of the Ricci tensor, 
which are geometrically well defined quantities.

A well known result relates the radial stability of the photon orbit and the 
properties of nearby timelike circular geodesics. Let $r_{\rm ph}$ be the radius of the photon orbit, and assume this 
orbit is not marginally stable.
From Rayleigh's stability criterion for circular geodesics
\cite{abramowiczKluzniak2005ApSS,letelier2003PRD} we have the following correspondence between the properties of photon orbits and of nearby
timelike circular geodesics:
If the photon orbit is radially unstable, timelike circular geodesics are allowed only for \mbox{$r>r_{\rm ph}$} and are radially unstable in a neighbourhood of $r_{\rm ph}$. 
If the photon orbit is radially stable, timelike circular geodesics are allowed only for \mbox{$r<r_{\rm ph}$} and are radially stable locally.
This behavior is seen in many spacetimes with photon orbits and no horizon 
\cite{puglieseQuevedoRuffini2011PRD, vieiraMarekEtal2014PRD, stuchlikHledik2002AcPSl, garciaHackmannEta2015JMP, boshkakayev2016PRD}.
Our findings demonstrate that the full linear stability analysis near stable photon orbits
relies heavily on the properties of off-equatorial motion and, by Eqs.~(\ref{eq:w2+w2}) and (\ref{eq:Rmunustrong}), on the local properties of 
the Ricci tensor. In general relativity, the presence of
a region of stability up to the stable photon orbit is therefore strongly dependent on the local properties of matter. 

It was shown in \cite{kluzniakRosinska2013MNRAS} that the sum \mbox{$\omega_r^2+\omega_\theta^2$} vanishes at the photon orbits of Kerr spacetime. 
The question of whether the formalism presented here extends to rotating, stationary
spacetimes remains an open problem; our results are a starting point to tackle this more general case. Based on the results for Kerr spacetime \cite{kluzniakRosinska2013MNRAS}, it is reasonable to suppose that the quantity $\omega_r^2+\omega_\theta^2$ might vanish at photon radii of stationary vacuum spacetimes (for both prograde and retrograde orbits), a hypothesis that deserves a thorough investigation. We also conjecture that the strong energy condition should be sufficient to ensure the positivity of $\omega_r^2+\omega_\theta^2$ for all circular timelike geodesics in stationary, axially symmetric spacetimes.

%
\acknowledgments
W.K. acknowledges support from the Polish NCN Grants No.~2013/08/A/ST9/00795 and No.~2013/10/M/ST9/00729. M.A. was supported by the Polish NCN Grant No.~2015/19/B/ST9/01099.
R.S.S.V. thanks the financial support from ``S\~ao Paulo Research Foundation'' (FAPESP), Grants No.~2010/00487-9, No.~2013/01001-0, and No.~2015/10577-9, and the hospitality of Nicolaus Copernicus Astronomical Center, where most of this research was developed.
%
%

\bibliography{refs_w2w2_arxiv}

\end{document}